\documentclass[10pt]{iopart}
\usepackage{epsfig}
\begin{document}
\title[Cosmology with coalescing massive black holes]
{Cosmology with coalescing massive black holes}

\author{Scott A Hughes\footnote[1]{Current address: Department
of Physics, Massachusetts Inst. of Technology, Cambridge
MA 02139} and Daniel E Holz\footnote[2]{Current address: Kavli Inst. for Cosmological
Physics, The Univ. of Chicago, Chicago, IL 60637}}
\address{Kavli Institute for Theoretical Physics, University of
California, Santa Barbara CA 93106}
\begin{abstract}
The gravitational waves generated in the coalescence of massive binary
black holes will be measurable by LISA to enormous
distances. Redshifts $z \sim 10$ or larger (depending somewhat on the
mass of the binary) can potentially be probed by such measurements,
suggesting that binary coalescences can be made into cosmological
tools.  We discuss two particularly interesting types of probes.
First, by combining gravitational-wave measurements with information
about the universe's cosmography, we can study the evolution of black
hole masses and merger rates as a function of redshift, providing
information about the growth of structures at high redshift and
possibly constraining hierarchical merger scenarios.  Second, {\it if}
it is possible to associate an ``electromagnetic'' counterpart with a
coalescence, it may be possible to measure both redshift and
luminosity distance to an event with less than $\sim 1\%$ error.  Such
a measurement would constitute an amazingly precise cosmological
standard candle.  Unfortunately, gravitational lensing uncertainties
will reduce the quality of this candle significantly.  Though not as
amazing as might have been hoped, such a candle would nonetheless very
usefully complement other distance-redshift probes, in particular
providing a valuable check on systematic effects in such measurements.
\end{abstract}


\vskip 2pc

As other contributions to these Proceedings will make clear, there are
currently major uncertainties in our understanding of the astrophysics
of massive binary black hole coalescences.  We are quite unsure how
often black holes merge, or what mass spectrum describes mergers, or
how these quantities are likely to evolve as the universe evolves.
What is certain is that if such mergers occur for binaries whose total
masses $M$ are roughly in the range $10^4\,M_\odot < (1 + z)M <
10^7\,M_\odot$, LISA will measure these waves out to redshifts of
order 10.  Even if these events are rare, LISA will measure them; and,
because the source of these waves arises at such large distances,
these measurements have the potential to provide detailed information
about the large scale structure of the universe.  In this
contribution, we discuss the potential of such measurements as
cosmological probes.

Coalescing binaries can be considered standard candles because general
relativity predicts a unique form for the two polarizations of the
binary waveform.  For LISA measurements, the strongest $l = 2$, $m =
2$ harmonic of the emitted waves has the form
\begin{eqnarray}
h_+ &=& {2 {\cal M}^{5/3}\over r}[\pi f(t)]^{2/3}\left[1 + ({\hat
L}\cdot{\hat n})^2\right]\cos\left[\Phi(t)\right]\;,
\nonumber\\
h_\times &=& {4 {\cal M}^{5/3}\over r}[\pi f(t)]^{2/3}\left[{\hat
L}\cdot{\hat n}\right]\sin\left[\Phi(t)\right]\;.
\label{eq:waveform}
\end{eqnarray}
In this equation, ${\cal M} = m_1^{3/5} m_2^{3/5}/(m_1 + m_2)^{1/5}$
is the so-called ``chirp mass''.  The vector $\hat L$ gives the
orientation of the binary (it is the direction of the orbital angular
momentum); $\hat n$ is the direction to the source according to an
observer that rides along with the LISA antenna.  The phase function
$\Phi(t)$ depends on intrinsic parameters such as masses and spins,
and so should be written $\Phi(t;m_1,m_2,{\vec S}_1,{\vec S}_2)$.  The
frequency $f(t) = d\Phi/dt$, and $r$ is the distance to the source.
Note that the waveform (\ref{eq:waveform}) does not include certain
effects such as Lense-Thirring precession of the binary's orbital
plane and multipoles other than $l = m = 2$; this will likely impact
the quantitative details of our results somewhat.

By accurately measuring the evolution of the phase function $\Phi(t)$,
we measure the binary's intrinsic parameters.  In particular, the
chirp mass ${\cal M}$ can be measured with exquisite precision:
${\delta \cal M/M}\sim 10^{-4}$ is a reasonable expectation.  This is
because ${\cal M}$ largely determines the total accumulated phase in
the measurement, and hence is very sensitive to that number; cf.\
Refs.\ {\cite{fc93,cf94,pw95}}.  The LISA antenna pattern and
modulations that arise from the antenna's orbital motion constrain
${\hat L}\cdot{\hat n}$ fairly well.  The only remaining parameter in
the waveform is the source distance $r$.  Most error in determining
$r$ comes from correlations with orientation and position errors
{\cite{c98}}; in practice, $r$ is likely to be measured to a precision
$\delta r/r \sim 1 - 25\%$ {\cite{h02}}.

Interpreting these formulas becomes somewhat more complicated when the
source generating $h_+$ and $h_\times$ is at a distance where
cosmological effects are important.  Without going into the details
(see {\cite{fh98,finn96}} for further discussion), Eq.\
(\ref{eq:waveform}) still works provided we redshift all frequencies
and timescales, and replace the naive distance measure $r$ with the
transverse comoving distance $D_M$ (see Ref.\ {\cite{hogg}} for a
detailed description of cosmological distance measures).  Putting
these two replacements into Eq.\ (\ref{eq:waveform}) and using the
fact that the luminosity distance $D_L = (1 + z)D_M$, we find
\begin{eqnarray}
h_+ &=& {2 {\left[(1 + z){\cal M}\right]}^{5/3}\over D_L}[\pi f(t)]^{2/3}
\left[1 + ({\hat L}\cdot{\hat
n})^2\right]\cos\left[\Phi(t)\right]\;,
\nonumber\\
h_\times &=& {4 {\left[(1 + z){\cal M}\right]}^{5/3}\over D_L}
[\pi f(t)]^{2/3}\left[{\hat L}\cdot{\hat
n}\right]\sin\left[\Phi(t)\right]\;.
\label{eq:waveform_z}
\end{eqnarray}
Now $\Phi(t)$ is strictly the {\it measured} gravitational-wave phase
function (and $f(t)$ is likewise the measured instantaneous
frequency).  It depends on {\it redshifted} values of the intrinsic
parameters: $\Phi(t) = \Phi[t; (1 + z)m_1, (1 + z)m_2, (1 + z)^2 {\vec
S}_1, (1 + z)^2 {\vec S}_2]$.  The reason for redshifting these
parameters can be simply explained using dimensional analysis.  In
general relativity, a mass $m$ can only impact the evolution of the
system as a timescale $\tau_m = G m/c^3$.  (General relativity has no
intrinsic scale, so the scales seen in any particular problem must
follow from that problem's specific parameters, such as masses.)  When
the system is placed at redshift $z$, the timescale is redshifted.
Thus, the apparent mass likewise picks up the factor $1 + z$.
Similarly, a spin $S$ impacts the system as a squared timescale
$\tau^2_S = G S/c^4$ and picks up a factor $(1 + z)^2$.  As a
consequence of this, the phase evolution of a cosmologically distant
binary is indistinguishable from a ``local'' binary with redshifted
masses and spins.

Measuring the waves from a distant binary black hole coalescence thus
provides us with two particularly interesting types of information ---
redshifted masses of the form $(1 + z)m$, and the luminosity distance
$D_L$ (as well as information about the source's location on the sky
and the spins of the binary's members).  There are two direct ways to
exploit this information for cosmological studies.  First, we can
assume that we know the universe's cosmography.  This allows us to
build a map between $z$ and $D_L$.  From our inferred $z(D_L)$, we can
break the mass-redshift degeneracy and learn about black hole masses
as a function of redshift.  Second, if it is somehow possible to
obtain the redshift independently, one can use the simultaneous
measurement of $z$ and $D_L$ to improve the cosmography.  (In
principle, there is also a third track: if one knows the mass spectrum
of the binaries, or has enough events to statistically sample the
range of the spectrum, then it should be possible to infer the
cosmological properties of this distribution; see Ref.\
{\cite{fc_cosmo}} for details.  We thank Sam Finn for pointing this
out to us.  Given the large uncertainties in our understanding of the
massive black hole merger rate and the likely breadth of the merger
mass spectrum, we will not discuss this third direction here.)

Let us begin by considering the first possibility.  We will describe
the universe using the currently popular ``concordance cosmology'',
with matter density given by $\Omega_m \simeq 0.35$, dark energy in
the form of a cosmological constant (that is, with equation of state
$p = w\rho$ and $w = -1$) with $\Omega_\Lambda \simeq 0.65$, and a
Hubble constant whose present value is $H_0 = 65\,{\rm km/sec}\,{\rm
Mpc}^{-1}$.  We will assume the universe is precisely flat, and that
the relative errors in $\Omega_\Lambda$ and $H_0$ are $\sim 10\%$
{\cite{wtz}}.  These assumptions allow us to build a map from redshift
to luminosity distance {\cite{hogg}} which is simple to invert (at
least numerically).  We then ask: How well can LISA measure the
parameters characterizing massive binary black hole systems,
particular the luminosity distance, redshift, and masses?  To do this
calculation, we use the restricted second post-Newtonian waveform
described in Ref.\ {\cite{pw95}} and a description of the LISA antenna
as described in Ref.\ {\cite{c98}}; see Ref.\ {\cite{h02}} for
complete details.

\begin{figure}
\begin{center}
\epsfig{file = 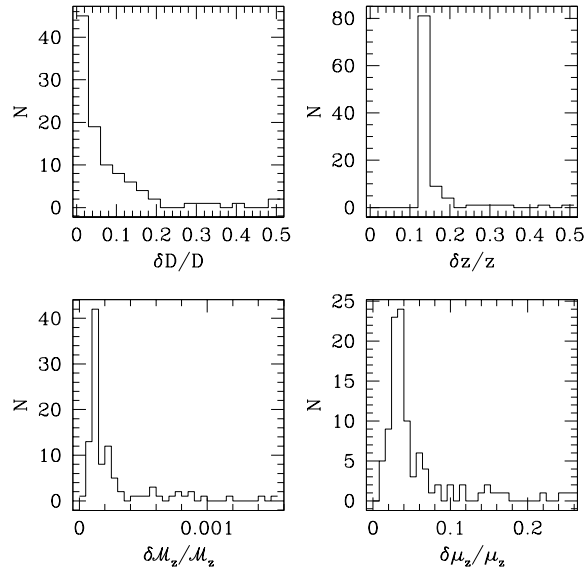, width = 8cm}
\caption{Parameter measurement accuracy distributions for binaries
with $m_1 = m_2 = 10^5\,M_\odot$ at $z = 1$.  In this case, the
parameters are measured particularly well: the peak determination of
luminosity distance is at $\delta D_L/D_L \simeq 2\%$, the peak in the
redshifted chirp mass is at $\delta {\cal M}_z/{\cal M}_z \simeq
10^{-4}$, and the peak in the redshifted reduced mass is at
$\delta\mu_z/\mu_z \simeq 4\%$.  The relatively large error in
redshift determination ($\delta z/z \simeq 15\%$) is because our
cosmological model assumes that the cosmological parameters are
themselves only accurate to about $10\%$.  If those parameters were
known precisely, we would find $\delta z/z \simeq \delta D_L/D_L$.}
\label{fig:z1_m1e5}
\end{center}
\end{figure}

An example of a binary that is measured particularly well is shown in
Fig.\ {\ref{fig:z1_m1e5}}.  The histograms describe parameter
measurement accuracies found by randomly distributing 100 binaries
over the sky at $z = 1$ with random orientations, and with the merger
time randomly distribed within an assumed 3 year LISA mission.  Each
binary is taken to have masses $m_1 = m_2 = 10^5\,M_\odot$.  In many
cases the redshifted masses and luminosity distance are measured with
exquisite precision.  The distribution for the distance peaks at
$\delta D_L/D_L \simeq 2\%$, and most events have $\delta D_L/D_L <
20\%$; the peak for the redshifted chirp mass is at $\delta {\cal
M}_z/{\cal M}_z \simeq 10^{-4}$, with most events having $\delta {\cal
M}_z/{\cal M}_z < 10^{-3}$; and the peak for the redshifted reduced
mass is at $\delta \mu_z/\mu_z \simeq 4\%$, with most events at
$\delta \mu_z/\mu_z < 10\%$.  The redshift does not appear to be
determined as well, but this is entirely due to our assumed errors in
the cosmography.  If the parameters determining the mapping between
$z$ and $D_L$ were known precisely, we would find $\delta z/z \simeq
\delta D_L/D_L$.  Again, we emphasize that these distributions are
computed with the restricted waveform given by Eq.\
(\ref{eq:waveform}), are likely to change when effects we have
neglected are taken into account.

The example shown in Fig.\ {\ref{fig:z1_m1e5}} is particularly good,
but is not far off what can be achieved for a broad range of system
masses in the rough band $10^4\,M_\odot < (1 + z)M_{\rm total} < ({\rm
several})\times10^6\,M_\odot$.  In this band, the redshifted masses
and the distance are typically measured with a relative error of a few
tens of percent.  These numbers are also approximately independent of
mass ratio, at least for $m_1/m_2 > 1/10$ or so --- the loss in
signal-to-noise ratio that comes from the reduced mass ratio is mostly
compensated by an increase in the number of measured cycles, so that
measurement precision remains roughly constant.

We thus conclude that, at the very least and using gravitational-wave
information alone, LISA will be able to provide useful and interesting
data on black hole masses and redshifts, making it possible to study
the mass and merger history of black holes in the universe.
Particularly at moderate to high redshift, this could provide a wealth
of data on the formation and evolution of the universe's structures
{\cite{h02,mhz01,menou}}.

Before moving on to the next track in our study --- redshift
determined independently --- we would like to comment on the accuracy
with which distances are determined.  Distance error is strongly
correlated with the errors with which the orientation and sky position
of the source are determined: the measured waveform (which is a
weighted sum of the two polarizations $h_+$ and $h_\times$) takes the
form
\begin{equation}
h_{\rm meas} = {[(1 + z){\cal M}]^{5/2}\over D_L}[\pi f]^{2/3}
{\cal F}({\rm angles})\cos\left[\Phi(t) + \phi({\rm angles})\right]\;.
\label{eq:wave_meas}
\end{equation}
The functions ${\cal F}({\rm angles})$ and $\phi({\rm angles})$
schematically indicate the dependence of the measured amplitude and
phase on a source's position and orientation angles; see {\cite{c98}}
for further discussion and details.  To measure the distance
accurately, we must nail down these angles as precisely as possible.

\begin{figure}
\begin{center}
\epsfig{file = 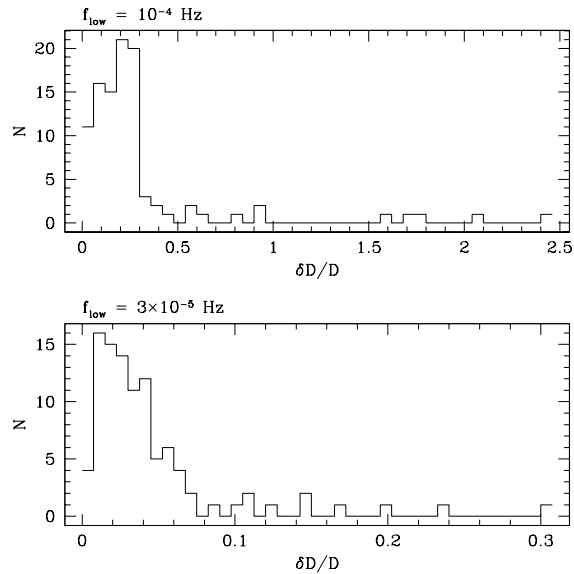, width = 8cm}
\caption{Comparison of distance determination for binaries at $z = 1$
with $m_1 = m_2 = 10^6\,M_\odot$.  In the top panel, we assume that
LISA's noise gets extremely bad below $10^{-4}\,{\rm Hz}$; the bottom
panel assumes the cutoff is at $3\times10^{-5}\,{\rm Hz}$.  Distance
determination improves by about a factor of ten in the lower panel.
This is because the wider band allows LISA to follow the binaries'
phase evolution for a much longer time: in the top panel, the binaries
radiate in band for about 15 days; in the bottom, the time in band is
about 10 months.  Controlling the low frequency performance will have
an important impact on LISA's science on high mass binaries.}
\label{fig:comp_noise}
\end{center}
\end{figure}

As has already been mentioned, the various source angles are
determined by exploiting LISA's orbital motion.  Roughly speaking, the
``angles'' indicated schematically in Eq.\ (\ref{eq:wave_meas}) are
effectively time dependent from the viewpoint of an observer who rides
along with the LISA antenna (though of course they are constant with
respect to the solar system's barycentre).  This time dependence
modulates $h_{\rm meas}$, with the exact modulation encoding the
values of the source angles.  As a rough rule of thumb, measuring the
angles well requires that LISA move through at least one radian of its
orbit, translating to a rough minimum of 2 months of observation to
adequately pin down the source angles.  Sources that don't radiate in
band for long enough (typically high mass systems) tend to determine
these angles badly and hence have poor distance determinations.  A
cure for these systems is to open the LISA band by reducing noise at
the low frequency end.  Figure {\ref{fig:comp_noise}} compares how
well the luminosity distance is determined under the assumption LISA's
noise becomes extremely large below $f = 10^{-4}\,{\rm Hz}$ (top
panel) and below $f = 3\times 10^{-5}\,{\rm Hz}$ (bottom panel).  Both
cases look at measurements of binaries that have $m_1 = m_2 =
10^6\,M_\odot$ and $z = 1$; as in Fig.\ {\ref{fig:z1_m1e5}}, we
randomly distribute the binaries' orientations, sky positions, and
merger times (within an assumed 3 year mission).  In the first case
($10^{-4}\,{\rm Hz}$ cutoff), the binaries only radiate in band for
about 15 days.  Distance determination is concomitantly poor, with a
peak in the distribution at $\delta D_L/D_L \simeq 25\%$.  (This
actually isn't {\it too} bad, largely because these binaries generate
a very strong signal.)  When the cutoff is lowered to
$3\times10^{-5}\,{\rm Hz}$, the binaries radiate in band for 10
months, and the distance determination is dramatically improved ---
the peak is now at $\delta D_L/D_L \simeq 2\%$.  Keeping the low
frequency behavior under control is clearly very desirable in order to
study high mass binaries.

\begin{figure}
\begin{center}
\epsfig{file = 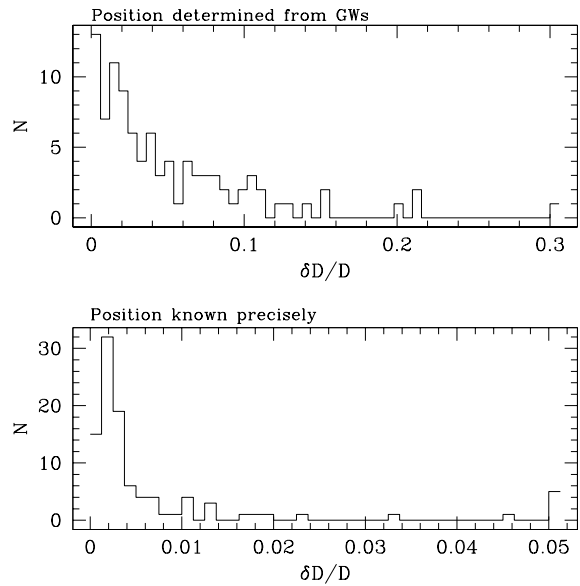, width = 8cm}
\caption{Comparison of distance determination for binaries at $z = 1$
with $m_1 = m_2 = 10^5\,M_\odot$.  The top panel shows the
distribution for determination with gravitational waves alone.  The
bottom panel assumes that an ``electromagnetic'' counterpart provides
an independent pointing solution for the binary.  This breaks many
degeneracies in the parameter determination, improving the distance
accuracy by about an order of magnitude (note the different scales in
the two panels).}
\label{fig:z1_m1e5_GWvsEM}
\end{center}
\end{figure}

Even in the best case, the position determination that LISA is likely
to achieve is not great by the usual standards of astronomy --- the
best angular resolution is on the order of several to several tens of
arcminutes. {\it If} it is possible to associate the gravitational
waves from a binary black hole merger with some kind of
``electromagnetic'' counterpart, the situation improves dramatically.
By getting an independent pointing solution for the binary, many of
the degeneracies that impact distance determination are broken.  We
illustrate this in Fig.\ {\ref{fig:z1_m1e5_GWvsEM}}.  The top panel
shows the distribution of $\delta D_L/D_L$ that can be expected when
only gravitational waves are used to analyze the binary.  The lower
panel shows how the distribution changes if a counterpart exists and
provides an independent pointing solution.  The distance precision is
improved by about an order of magnitude in this case.

Associating a counterpart with a merger is not going to be easy.  The
number of galaxies in each LISA ``pixel'' can reasonably be expected
to number in the hundreds.  Also, it is far from clear what kind of
electromagnetic signature will characterize a counterpart.  Some work
discussing the kinds of counterparts one can imagine has appeared
{\cite{shvsb88,lv96,an02}}; we hope that the promise of coordinating
electromagnetic observations with gravitational-wave measurements will
motivate additional work in this vein.

Associating a counterpart with a gravitational-wave merger measurement
offers another exciting possibility: from such an association, it may
be possible to directly measure the event's redshift, rather than
inferring it by combining the distance with cosmographic information
{\cite{bernie}}.  In principle, this could provide simultaneous
measurements of $D_L$ and $z$, each with precision $<1\%$.  Such a
measurement, though possibly rare and difficult to make, would provide
invaluable information about our cosmography (complementing other well
developed probes such as Type-Ia supernovae {\cite{p99,r98}}).  At the
very least, because the systematics are so different from that of
other candles, a merger would increase confidence in all candles
(assuming that their measurements are in accord!).  At best, because
the merger candle could be exquisitely precise, it could impact
cosmological parameter determination with high weight.

\begin{figure}
\begin{center}
\epsfig{file = 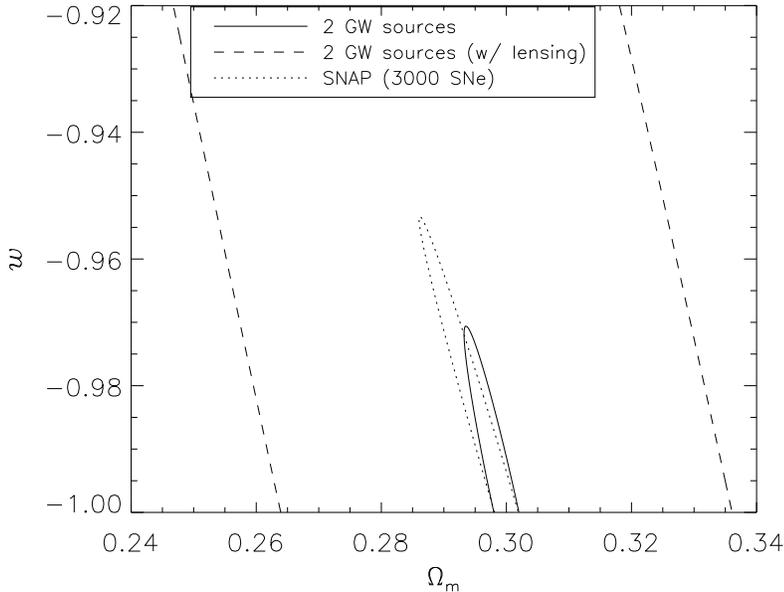, width = 11cm}
\caption{Likelihood contours (1 sigma) for the matter density
$\Omega_m$ and the equation of state parameter $w$ (relating the
pressure and density of dark energy, $p = w\rho$).  We assume that the
universe is flat, and that the underlying cosmology has $\Omega_m =
0.3$, $w = -1$.  We compare parameter determination following
measurement of two gravitational-wave candles (at $z = 1$ and $z =
3$), and 3000 Type-Ia supernovae with the SNAP {\cite{snap}} (evenly
distributed from $z = 0.7$ to $z = 1.7$).  Gravitational lensing has a
dramatic impact on the gravitational-wave candle, puffing the
likelihood contour out by a large amount.  The event rate of
supernovae is high enough that lensing effects can be averaged out.}
\label{fig:contour}
\end{center}
\end{figure}

In fact, as has recently been shown {\cite{whm02}}, {\it all} candles
have a fundamental limit to their precision set by gravitational
lensing, arising from the propagation of radiation through the lumpy,
inhomogeneous universe in which we live.  Lensing impacts
gravitational waves exactly as it impacts electromagnetic radiation.
A lens with magnification $\mu$ will cause an event whose true
luminosity distance is $D_L$ to appear to be at distance
$D_L/\sqrt{\mu}$.  (Note that $\mu$ can be less than 1 ---
``demagnification'' is in fact quite likely.)  By convolving this
error with the expected magnification distribution $P(\mu)$
{\cite{whm02}}, we find that $\delta D_L/D_L \simeq 5 - 10\%$ due to
lensing is probable (with some dependence on the redshift of the
merger event) {\cite{hh_inprep}}.  Although intrinsically of high
quality, the actual effectiveness of the merger plus counterpart
standard candle will be significantly reduced by lensing.

Figure {\ref{fig:contour}} illustrates cosmological parameter
determination using standard candles.  We assume that the universe is
flat, and that the total density is given by matter ($\Omega_m$) plus
``dark energy'' with equation of state $p = w\rho$.  The figure shows
the 1-$\sigma$ likelihood contours in the $\Omega_m$-$w$ plane for
several cases.  The heavy black line shows the contour that would be
obtained if two merger events with counterparts are measured,
neglecting gravitational lensing.  This contour is nearly identical to
(indeed, slightly tighter than) that expected for 3000 Type-Ia
supernovae measured by the proposed SNAP satellite (dotted line)
{\cite{snap}}.  The dashed line illustrates what happens to the heavy
black line when the systematic uncertainty induced by gravitational
lensing is taken into account.  It's rather sobering to note how much
of an effect the lensing has on the gravitational-wave measurements.
Lensing does not impact supernovae nearly as much: the supernova event
rate is high enough that measurements can average away lensing
effects, essentially sampling the full range of the lensing
probability distribution.

Combining these lensed gravitational wave events with the supernovae
changes the supernova contour very little.  We believe that, in the
end, the most important contribution of a gravitational-wave standard
candle will be as a check on systematic and evolutionary effects in
the candle dataset.  As a candle with drastically different
properties, coordinated gravitational wave/electromagnetic
measurements would improve confidence in all standard candles.

\section*{Acknowledgments}

We thank Shane Larson, David Merritt, Kristen Menou, and Steinn
Sigurdsson for useful discussions.  This work was support by NSF Grant
PHY--9907949.

\end{document}